# Deep-Learning-Based Prediction of Nanoparticle Phase Transitions During In Situ Transmission Electron Microscopy

**Authors:** Wenkai Fu[1], Steven R. Spurgeon[1,2], Chongmin Wang[1,§], Yuyan Shao[1], Wei Wang[1,‡], Amra Peles[1,*]

**Affiliations:**
1. Pacific Northwest National Laboratory, 902 Battelle Blvd, Richland, WA 99354, USA
2. Department of Physics, University of Washington, Seattle, WA 98195, USA

\* Corresponding author. Email: amra.peles@ornl.gov

§ Corresponding author. Email: chongmin.wang@pnnl.gov

‡ Corresponding author. Email: wei.wang@pnnl.gov

**Abstract:** We develop the machine learning capability to predict a time sequence of in-situ transmission electron microscopy (TEM) video frames based on the combined long-short-term-memory (LSTM) algorithm and the features de-entanglement method. We train deep learning models to predict a sequence of future video frames based on the input of a sequence of previous frames. This unique capability provides insight into size dependent structural changes in Au nanoparticles under dynamic reaction condition using in-situ environmental TEM data, informing models of morphological evolution and catalytic properties. The model performance and achieved accuracy of predictions are desirable based on, for scientific data characteristic, based on limited size of training data sets. The model convergence and values for the loss function mean square error show dependence on the training strategy, and structural similarity measure between predicted structure images and ground truth reaches the value of about 0.7. This computed structural similarity is smaller than values obtained when the deep learning architecture is trained using much larger benchmark data sets, it is sufficient to show the structural transition of Au nanoparticles. While performance parameters of our model applied to scientific data fall short of those achieved for the non-scientific big data sets, we demonstrate model ability to predict the evolution, even including the particle structural phase transformation, of Au nano particles as catalyst for CO oxidation under the chemical reaction conditions. Using this approach, it may be possible to anticipate the next steps of a chemical reaction for emerging automated experimentation platforms.

# 1   Introduction

The aim of video forecasting is to predict the future sequence of video frames based on a previous sequence of frames, and deep learning has been used to predict videos evolution in a wide range of fields, such as weather forecasting [1], autonomous driving [2], and robotics [3]. In these video prediction tasks, deep convolutional neural network methods are typically used to learn features from images, and then recurrent neural networks methods are used to predict the evolution of a sequence of such features [4]. Mathematically, a spatiotemporal video prediction gives the most probable length-$K$ sequence of frames in the future from a time $t$ based on the previous length-$J$ sequence of frames, i.e.,

$$\tilde{\chi}_{t+1}, \dots, \tilde{\chi}_{t+K} = argmax_{\chi_{t+1},\dots\chi_{t+K}} p(\chi_{t+1}, \dots, \chi_{t+K} | \chi_{t-J+1}, \dots, \chi_t). \quad (1)$$

Here $\chi \in \mathbb{R}^{P \times H \times W}$, is a tensor that represents a video frame with channels $P$, height $H$, and width $W$. For a grayscale image, $P$ is one, and the intensity varies between 0 and 255. For the spatiotemporal prediction as video forecasting, it is crucial for the model to memorize spatial appearances and temporal variations [5], and machine learning models based on convolutional neural networks (CNNs) and recurrent neural networks (RNNs) have been proposed. A seminal model for video prediction is the Convolutional Long Short-Term Memory (*ConvLSTM*), first proposed for precipitation nowcasting [1]. ConvLSTM overcomes the limitation of standard LSTM (a special recurrent neural network structure), where the spatial information is lost. In ConvLSTM, all the internal variables are with the same dimension as the input frame, i.e., 3D tensor of $\mathbb{R}^{P \times H \times W}$. Inspired by convLSTM, several more advanced deep learning models have been proposed recently, such as PredRNN [5] and PhyDNet [6]. PredRNN improves upon convLSTM by using a unified spatiotemporal memory pool. In this approach, memory is shared by all LSTM cells, whereas in convLSTM, memory is constrained in each LSTM layer. PhyDNet uses a two-branch deep architecture to disentangle physical dynamics from unknown complementary information. In one branch, a new physics-informed RNN cell named PhyCell is used to capture the physical dynamics, e.g., trajectory or spatial movement that can be described by partial differential equations (PDEs). In the second branch, the data-driven convLSTM is used to learn the unknown factors, such as appearance and texture. Then, the outputs of PhyCell and convLSTM are combined for a full prediction of the next frame [6]. Currently, PhyDNet is the state-of-the-art architecture for video prediction [6].

While the deep learning models have achieved good performances in standard benchmark datasets such as Moving MNIST [7], it is critical to apply these models on scientific data such as videos that are captured using in-situ transmission electron microscopy (TEM) to gain scientific insights [8, 9] into physical and chemical processes such as catalytic behavior under changing environments. Environmental in-situ TEM is a method that is used to directly observe structural evolution of materials under relevant device operating conditions. The mostly common method for capturing such a structural evolution is recording video, which typically offers a temporal resolution ranging from second to millisecond using a conventional charge coupled device (CCD) cameras. Although such a temporal resolution is slower than typical chemical reaction timelines of ~femtoseconds, the captured video is rich in information that is beyond the traditional way of a frame-by-frame image analysis [10, 11]. As a typical use case, we consider gold nanoparticles, which are a well-known catalyst for CO oxidation. The activity of gold nanoparticles shows an apparent size effect; that is, when the particle size is smaller than several nanometers, it shows extremely high activity unlike inert bulk gold. It is well known that the size, shape, and availability of reaction sites are key parameter influencing the catalytic activity [33]. The reaction environment becomes important, especially for real world applications where catalysis proceed under device operating conditions. He et al [9] have reported the shape change of Au nanoparticles during oxidation of CO using environmental TEM chamber by collecting in-situ TEM series of video frames. Two fundamental questions that need to be answered are: (1) where is the active site of the gold particle, and (2) how does the active particle become deactivated. By using environmental in-situ TEM, Chongmin et al have captured the structural evolution of gold nanoparticles of different size under dynamic catalytic reaction conditions [9]. It was reported that when the size of the gold nanoparticles is less than 4 nm, an exposure to CO causes continuous structural evolutions and transformation from the usual FCC structure; while, if the size is large than 4 nm, it retains an FCC structure, but with surface layer being slightly expanded. The atomic level dynamic evolution of gold under working condition carries essential information in terms of where is the active site and how they transform from active state into deactivated one. Hence, in this work, we aim to use machine learning technique to delineate details with temporal resolution.

Applying machine learning techniques on electron microscopy data is gaining interest and growing fast. Typical approaches have improved image segmentation capabilities to extract microstructural features from TEM images or video frames, beyond well-known image analysis technique [8, 10-15]. To name a few, the U-Net neural network was used for image segmentation task by identifying boundaries of nanoparticles in each frame of liquid-phase TEM videos [8], and few-shot learning approaches have recently been successfully applied to segment scanning TEM images [12]. Recently, an unsupervised algorithm named AutoDetect-mNP was reported for analysis of TEM images of metal nanoparticles (mNP) [13]. This algorithm can realize individual mNP detection, feature extraction of particle shape attributes, the resolution and filtering out of overlapping particles, and shape classification and statistical analysis. The performance of this algorithm has been demonstrated using Au, palladium, and cadmium selenide nano systems. Most of the data science research in electron microscopy data so far have focused on the single image analysis, image classification and features extraction.

In this work, we apply state of the art video prediction deep learning methods [6, 16-19] on the scientific, in-situ TEM time series of images to investigate the predictive ability of the nano-particles structural changes experimentally observed in [9]. We trained deep learning models to predict a sequence of future frames based on the input of a sequence of previous frames, i.e., a sequence-to-sequence task [17]. Because the TEM video frames used in this work recorded the structure transition of ultra-small gold nanoparticle from rigid to dynamic in its catalytic working environment [9], an important question to be answered by this study is whether the deep learning architecture can predict such a structural transition. The machine learning capability to predict the microstructural transition has many promising applications, including understanding nonequilibrium processes in chemical and materials systems, as well as the design of data-driven next-generation TEM instrumentation [20].

## 2 Specific scope of the study

Supported gold (Au) nanoparticles (NPs) show superior catalytic activity in the size of a few nanometer, but its origin is unknown for a long time [21-24]. Until recently, in-situ environmental TEM video revealed the evidence that this size effect of Au NP might be explained by the transition of its atomic structure [9]. Specifically, in the catalytic working condition, structure of 4-nm Au NP remains rigid and ordered; however, a structure of ultrasmall Au clusters, i.e., smaller than 2 nm and single layer, becomes disordered, or dynamic structure (DS) is formed. This structure transition of ultrasmall Au NP gives insight of the size effect on Au NP catalytic performance. Here we evaluate the performance of PhyDNet on in-situ TEM videos of Au NP, with the aim to examine the order-disorder phase transitions in these materials.

## 3 Methods

### 3.1 Architecture of PhyDNet

In this work, the state-of-the-art PhyDNet deep learning model was selected to predict in-situ TEM videos [6], and its architecture is shown in Figure 1. The model consists of an encoder, decoder, PhyCell, and convLSTM blocks, the last two of which are RNNs. To predict the frame at time $t + 1$ ($\chi_{t+1}$) based on the frame at time $t$ ($\chi_t$), $\chi_t$ is fed into the encoder that consists of six convolutional layers to extract features from the image, and a Leaky version of a Rectified Linear Unit (LeakyReLU) is following each convolutional layer as the activation function. Here, $\chi$ is the tensor that represents a video frame generated by reading the image into the code, as defined in the Introduction section. The output of the encode, $E(\chi_t)$, is fed into the two RNN neural networks to update their internal states. In PhyCell, $E(\chi_t)$ is fed into the two connected convolutional layers to generate the output labeled as $F[E(\chi_t)]$, then, $\widetilde{H}$ is computed by

$$\widetilde{H} = H_t^p + F[E(\chi_t)], \tag{2}$$

where $H_t^p$ is the hidden state at time $t$. Then, $E(\chi_t)$ and $\widetilde{H}$ are concatenated along the first axis, i.e., channel number, to form a tensor $J$. Then, $J$ is fed into the convgate layer, and a sigmoid function is applied to the output of the convgate layer to compute the tensor $K$. Finally, $H_{t+1}^p$, i.e., the hidden state at the next time step $t + 1$, is calculated by

$$H_{t+1}^p = \widetilde{H} + K \odot \left(E(\chi_t) - \widetilde{H}\right), \tag{3}$$

where $\odot$ is the Hadamard product.

In the convLSTM RNN, $E(\chi_t)$ is concatenated with the hidden state of the first convolutional layer at time $t$, $H_{0,t}^r$, along the first axis. The resulted concatenation is fed into the first convolution layer to calculate the output $X$. Then, $X$ is split along the first axis into four parts, and a sigmoid activation function is applied to these four parts to calculate four tensors named $i, f, o, g$. Then, cell output $C_{0,t+1}$ and $H_{0,t+1}$ is computed by

$$C_{0,t+1} = f \odot C_{0,t+1} + i \odot g, \tag{4}$$

$$H_{0,t+1} = o \odot \tanh(C_{0,t+1}). \tag{5}$$

The operation in the following convolution layers in convLSTM are similar except the input is a concatenation of $H_{i-1,t+1}^r$ and $H_{i,t}^r$ for layer $i$.

After $H_{t+1}^p$ and $H_{2,t+1}^r$ have been computed, they are added element-wise to calculate $H_{t+1}$, which is fed into the decoder to predict the next frame $\chi_{t+1}$.

The two-branch architecture of PhyCell and convLSTM in PhyDNet were designed to disentangle physical dynamics that can be described by partial differential equations (PDEs) from unphysical factors [6]. The physical dynamics are captured by PhyCell, which models a generic class of linear PDEs. The unphysical factors are simulated by the data-driven convLSTM. It should be noted that PhyDNet is a general architecture for video prediction. It has been applied to predict benchmark videos of moving digits [7], sea surface temperature [25], and moving human actions [26]. In this work, we used PhyDNet for the in-situ TEM video. In future work, PhyDNet can also be applied on other scientific videos, such as those from optical or atomic force microscopy.

The parameters inside the PhyDNet model were determined via the training process using the backpropagation algorithm, which is merely a practical application of the chain rule for derivatives. The parameters were determined to minimize the loss function of mean squared error. The Adam optimizer with learning rate of 0.0001 and 1000 epochs were used to train the PhyDNet model. The data to train the PhyDNet model were detailed in the following section, where each instance of data was a sequence of video frames extracted from the original TEM videos, or a video clip.

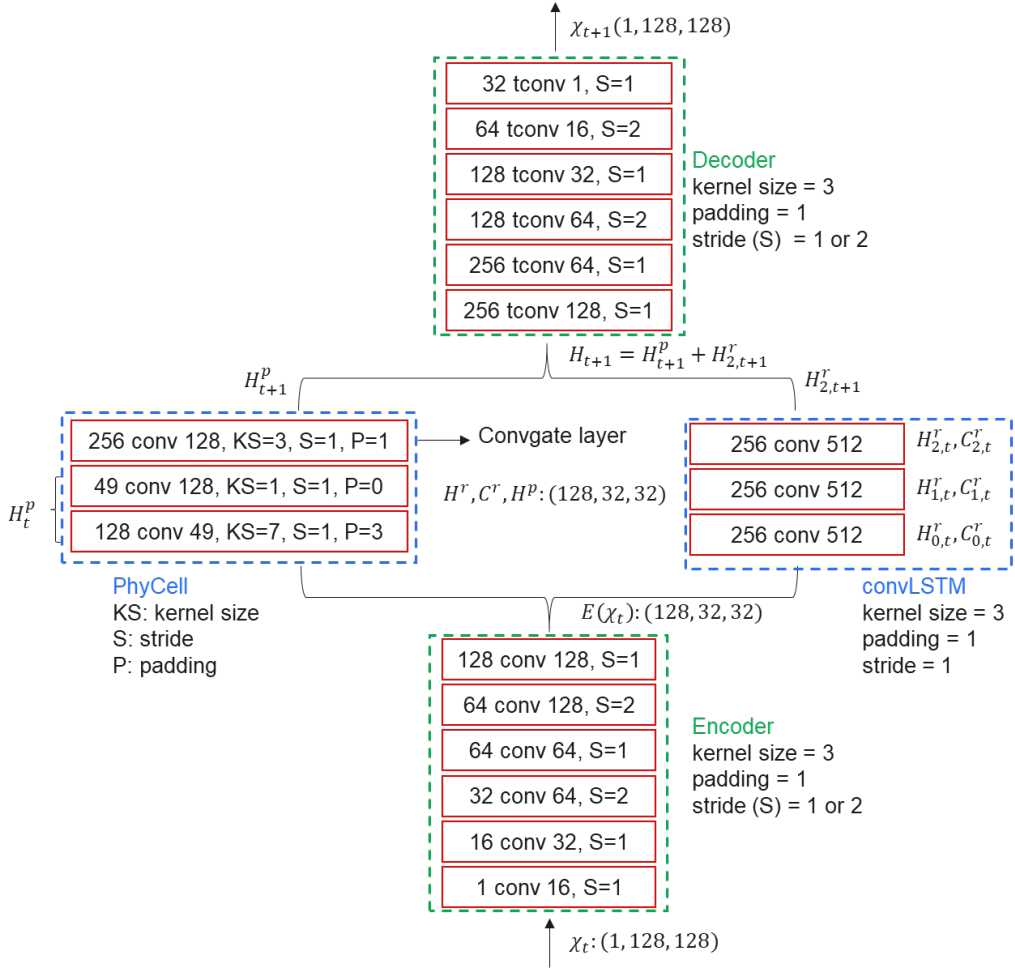

Figure 1: Architecture of the PhyDNet model. The input and output channels of the convolution layers are shown in the red boxes. Sizes of the tensors as $\chi_t$ and $E(\chi_t)$ are shown in their right. $H^p$ is the internal state of the PhyCell, and $H^r$ and $C^r$ are the internal states of the convLSTM.

## 3.2 Data preparation

To prepare data to train the deep learning models, frames were first extracted from the videos, which resulted in 551, 549, and 158 grayscale frames from the videos of 4-nm, 2-nm, and single layer (SL) Au NP, respectively. The original frames extracted from videos were with size 480x480 and show the gold nanoparticles supported on the $CeO_2$ surface [9], as listed in Figure S1(a-c). Due to the movement of the camera in the experiments, in these original frames, the relative locations of Au NP within the frames are not constant. Because the focus of this work is the structural dynamics of Au NP, the large CeO$_2$ and gas areas are useless. To clean the data and to reduce the computational load of the training, for each frame, a 128 x128 window was placed on the original frame manually such that the Au nanoparticle is at the center of the window. The cropped frames were normalized with zero mean and standard deviation of unity and used for the model. Figure S1(d-f) show examples of the cropped frames.

The processed frames were used to construct video clips for the training of the deep learning model, which is shown in Figure 2. Different training strategies identified by the number of frames ($N_f$) and number of steps ($N_s$) were tested. For each training strategy, one PhyDNet model was trained. $N_f$ sequential frames were fed into a PhyDNet model to predict the following $N_f$ sequential frames. Two adjacent frames in the $N_f$ sequential frames span $N_s$ steps in the original frames from the TEM video. The meanings of $N_f$ and $N_s$ are illustrated in Figure 2. The first frame of the $N_f$ sequence of frames was shifted by one frame to construct another video clip, as shown using the $N_f = 4$ and $N_s = 1$ and 2 cases in Figure 2. Following this data preparation, the numbers of resulted video clips for the 4-nm, 2-nm, and single layer Au nanoparticles are shown in Figure S2. Because the number of frames from the single-layer video is less than the number of frames from the 4-nm and 2-nm videos, only $N_f$ of 4 and 6 were considered. For each training strategy, the video clips were randomly split into a training set (90%) and a testing set (10%).

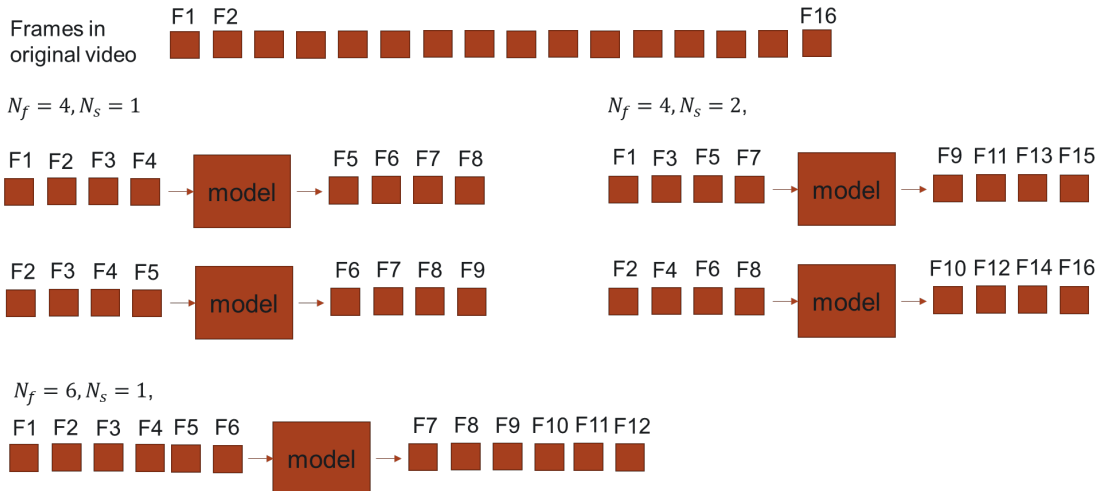

*Figure 2: Different strategies of constructing video clips from original TEM video frames. These strategies differ by number of frames in a video clip ($N_f$) and number of steps between two frames in the video clip ($N_s$). The resulted video clips of a strategy are used to train and test a PhyDNet model.*

It can be seen from Figure S2 that the number of video clips for 4-nm and 2-nm Au NPs are less than 550, and the number for SL Au NP video is even less. The volume of data is much less than standard benchmark datasets like Moving MNIST [7] and Human 3.6 [26], which contain more than tens of thousands or even millions of data sets for training and testing. The small amount of experimental data is typical since it usually takes lots of efforts to measure or record such scientific data. Hence, to apply a machine learning model to limited experimental data, it is important for such model to be data-efficient, i.e., the model needs to have good performance given limited amount of data, which requires that the number of training parameters in the model cannot be too large. The selected PhyDNet model meets this requirement because one of its recurrent branches, PhyCell, is physics-informed, which contains much less parameters than another data-driven recurrent branch, convLSTM. In our study, the number of training parameters in PhyCell is 609073, and the number of training parameters in convLSTM is 3540480. While with less training parameters, PhyCell can give better performance than the data-driven convLSTM, as reported in an ablation study of PhyDNet [6]. Hence, PhyDNet is suitable for scientific machine learning tasks where data is limited.

The predicted frames were histogram matched [27] with respect to the target frames to decrease the contrast difference between the two, which generated a new histogram matched predicted frames. As shown in Figure 3 and Figure 4 in the following section, the contrast decrease by histogram match algorithm does not affect the evaluation of the structural changes of the gold nanoparticle. It primarily improves the similarity in the gas and the $CeO_2$ support regions between the histogram-matched predicted and the target frames. Therefore, the metrics such as MSE and structural similarity (SSIM) between the histogram-matched predicted images and the target frames are better than using the original predicted images directly from PhyDNet models. In the following, these parameters were reported using the histogram matched predicted frames and the target frames.

## 4 Results and Discussion

### 4.1 Onset of dynamic structure

An important performance of PhyDNet is whether it can predict the transition of the Au NP from rigid to dynamic structure. Shown in Figure 3 and Figure 4 are the *testing* video clips of the PhyDNet predictions that contain the onset frames of the dynamic structures of the 2-nm and the single layer Au NPs, respectively. In Figure 3, frame 4 is the onset frame of the dynamic structure of the 2-nm gold nanoparticle, and in Figure 4, frame 2 is the onset frame of the dynamic structure of the single-layer gold nanoparticle. These onset frames can be easily determined manually from the original TEM videos. The results in Figure 3 and Figure 4 clearly show that the PhyDNet models predict the transitions of the 2-nm and single layer Au NPs from rigid to dynamic.

In addition, the effects of the histogram-match algorithm to remove the contrast difference between the predicted and target frames are clearly shown in Figure 3 and Figure 4. For instance, in the column of frame 4 in Figure 3, the morphology of the gold nanoparticle shown in the histogram-matched predicted frame (the second row) is similar with the original predicted frame (the third row). The histogram-match algorithm primarily makes the large areas of $CeO_2$ support and gas between the predicted and the target frames more similar.

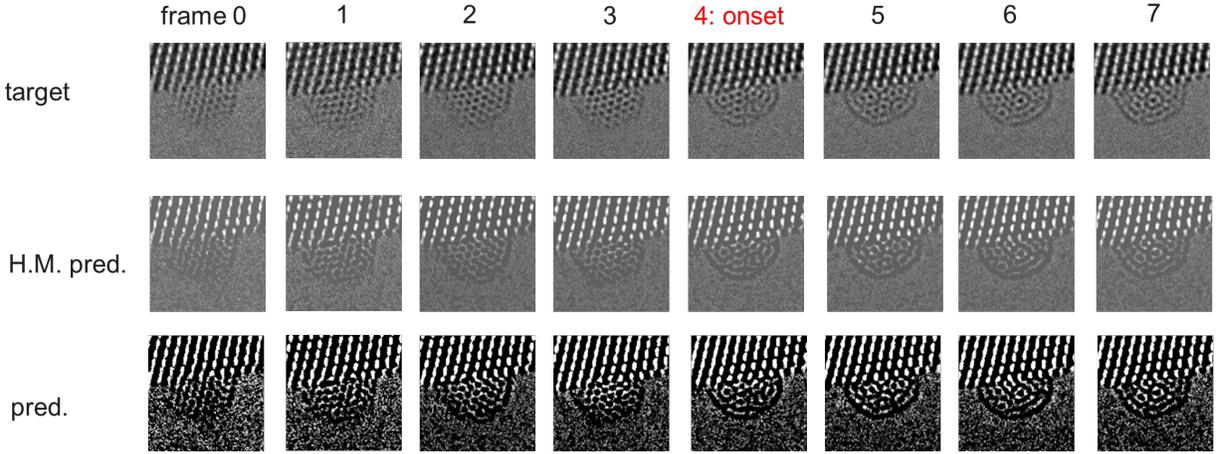

*Figure 3: A testing case that shows the transition of the 2-nm Au NP from rigid to dynamic. The onset of dynamic structure is at frame 4. The target, histogram matched (H.M.) predicted, and the predicted frames are listed. The machine learning model was trained using $N_f = 8, N_s = 8$, and epoch = 1000.*

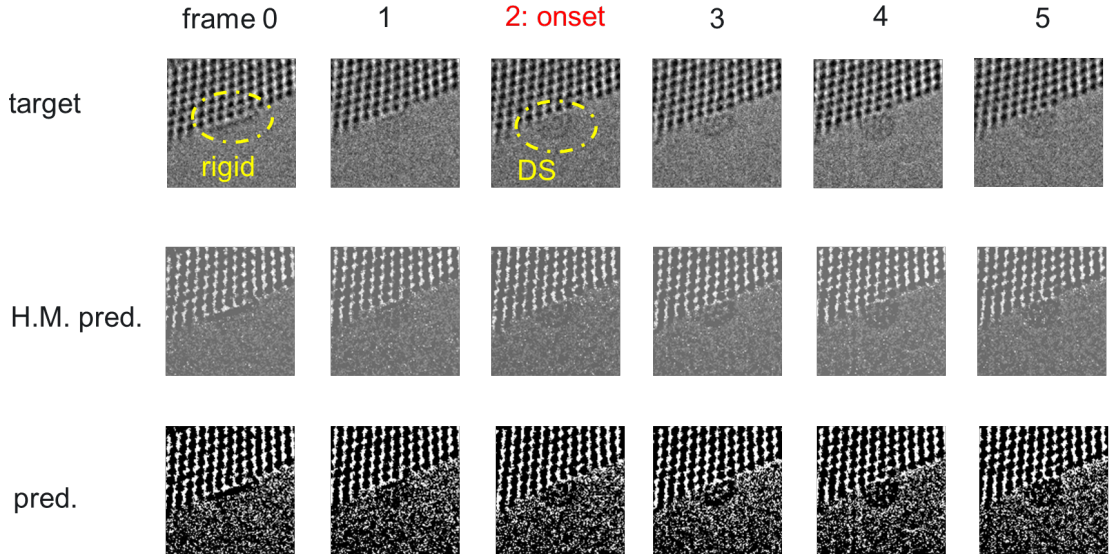

*Figure 4: A testing case that shows the transition of the single layer Au NP from rigid to dynamic. The onset of dynamic structure is at frame 2. The target, histogram matched (H.M.) predicted, and the predicted frames are listed. The deep learning model was trained using $N_f = 6, N_s = 2$, and epoch = 1000.*

## 4.2 MSE and SSIM of training strategy

While the results in Figure 3 and Figure 4 show qualitative similarity between the histogram matched predicted and target frames, to quantify such similarity, the mean square error (MSE) and the more advanced metrics of structural similarity (SSIM) [28] were calculated. The SSIM is a float number between zero and unity computed from two images. SSIM is zero for two totally different images and one for two identical images. The MSE and SSIM of different training strategies for the 4-nm, 2-nm, and single layer Au NP are shown in Figure 5, Figure 6, and Figure 7, respectively. For each training strategy, the metrics were averaged over all the video clips in the training, testing, and both sets, where each video clip contained $N_f$ pairs of target and predicted frames. The models trained after 1000 epochs were used to predict the frames.

From these results, it can be seen the SSIMs for majority of the training strategies are between 0.6 to 0.7. MSE and SSIM of the test cases are similar to that of the training cases. In general, with the increase of $N_f$, MSE decreases and SSIM increases, i.e., the prediction improves. The effect of $N_s$ on these metrics is not significant.

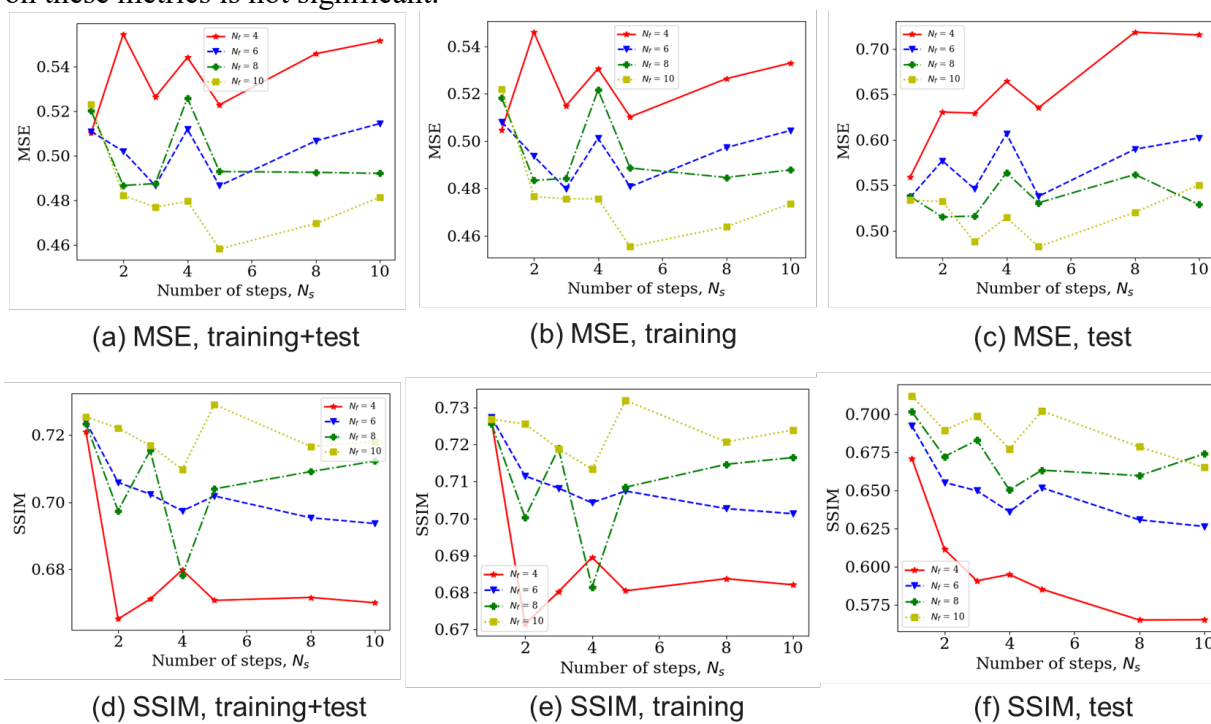

Figure 5: MSE and SSIM of the 4-nm Au NP as a function of training strategy.

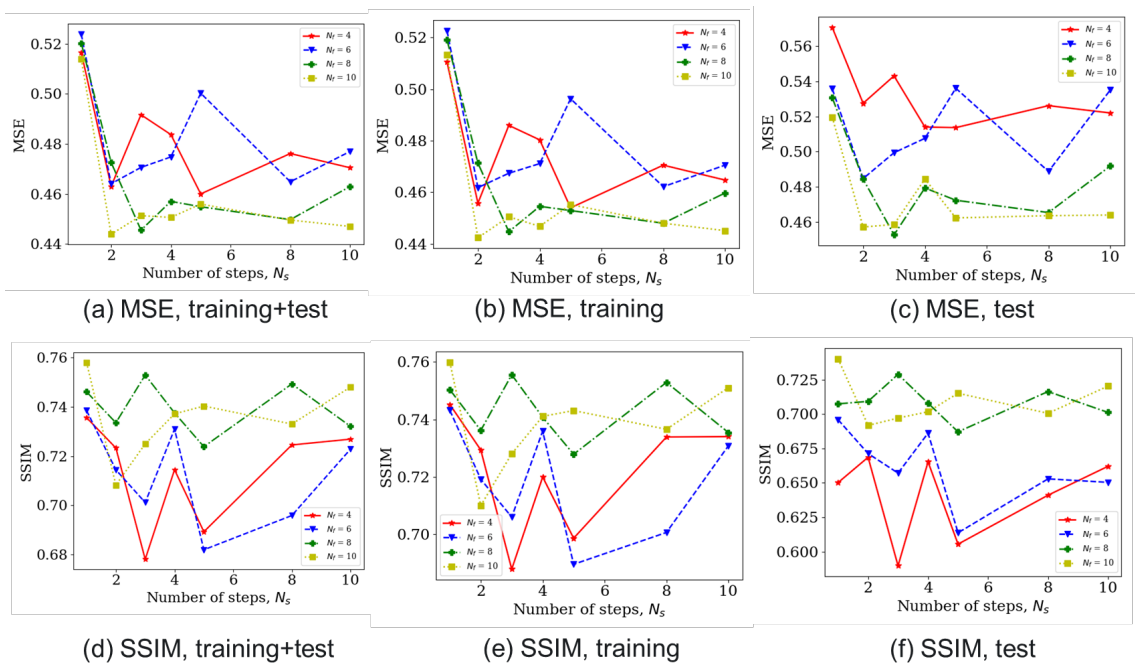

Figure 6: MSE and SSIM of the 2-nm Au NP as a function of training strategy.

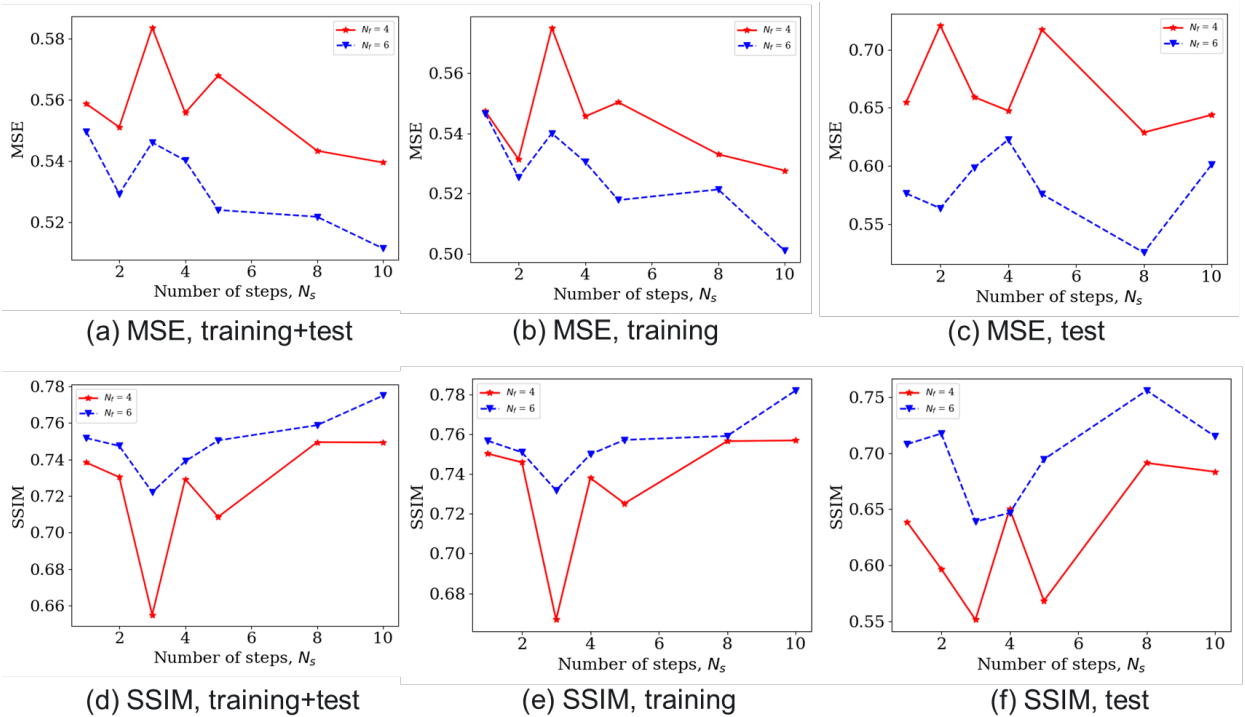

*Figure 7: MSE and SSIM of the single layer Au NP as a function of training strategy.*

## 4.3 MSE and SSIM at different frames

An important performance of the deep leaning model for video prediction is the predicted quality of each frame in the predicted video sequence. Shown in Figure 8 are the MSE and SSIM of each predicted frame in the training strategy of 2-nm video, $N_f$ of 6, and $N_s$ of 3. The metrics of a frame shown in Figure 8 are the average over all the video clips in the training, testing, and both sets. Prediction quality gradually decreases with the predicted further frames. For the testing set, the MSE increases from about 0.49 for the first predicted frame to 0.51 for the last predicted frame. The SSIM decreases from about 0.7 for the first predicted frame to about 0.62 for the last predicted frame. These results show the relatively stable prediction quality of PhyDNet over the frames in the video clips, which is also observed when PhyDNet was applied to standard datasets such as Moving MNIST, Traffic BJ, Sea Surface Temperature, and Human 3.6 [6].

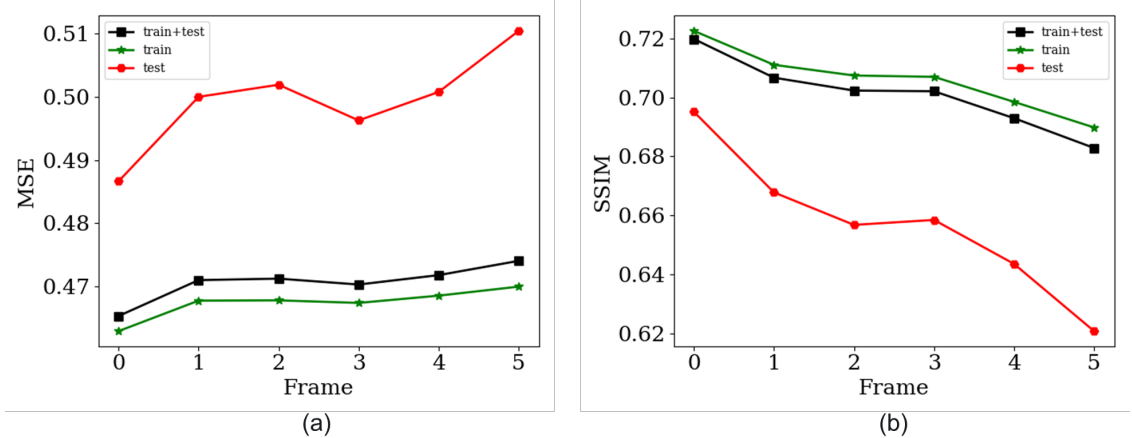

*Figure 8: MSE (a) and SSIM (b) of each predicted frame averaged over all the video clips in the training and testing, training, and testing datasets. The training strategy is 2-nm Au NP video, $N_f$ is 6, and $N_s$ is 3.*

## 4.4 Predictions using PhyDNet trained at different epochs

Figure 9 shows the predictions of a frame in a test case, and the training strategy is 2-nm Au NP video with $N_f = 6$ and $N_s = 3$. The models trained after different number of epochs were used to generate the predictions. It shows that the prediction performance improves with epoch. The predictions using models trained using larger number of epochs can capture the white noise in the gas region of the frame. The predicted structure of the gold nanoparticle becomes more similar to the target. For the same strategy, the computed SSIMs as a function of frame and epoch, as well as a function of epoch, training, and testing sets are shown in Figure 10. The performance improvement with epoch is clearly shown. The SSIM improves more significantly for epochs less than 600.

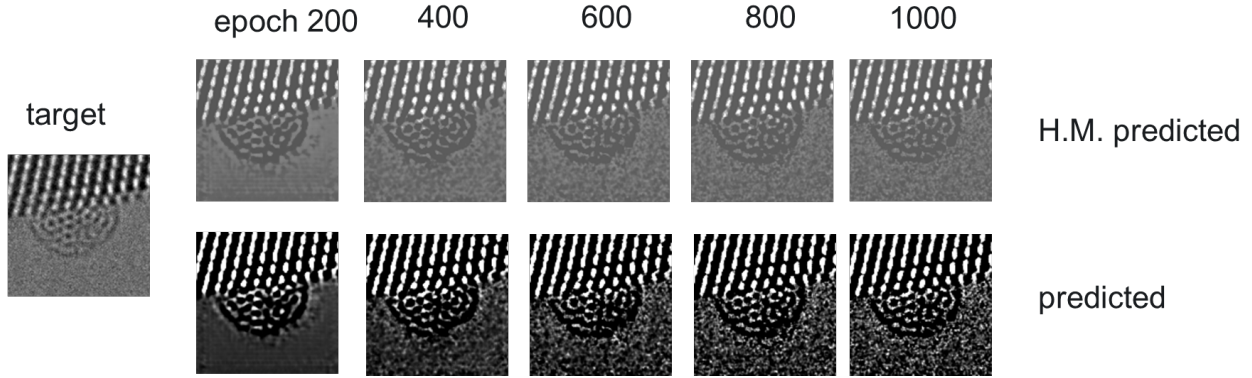

Figure 9: For the prediction of 2-nm Au NP video using $N_f = 6$ and $N_s = 3$, models trained after different number of epochs are used to predict onset frame of dynamic structure. The prediction improves with epoch. The predicted and histogram matched predicted frames are shown.

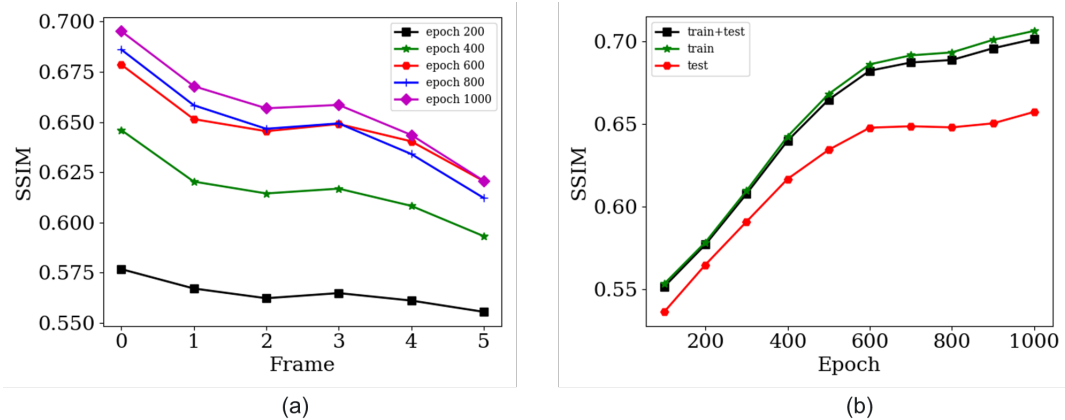

Figure 10: (a) SSIM as a function of frame predicted using models trained at different epochs. The SSIM of a frame is the average over all the testing instances. (b) SSIM averaged over all frames in training and testing, training, and testing instances using predictions of models trained at different epochs. In both subfigures, the training strategy is for the 2-nm Au NP video, $N_f = 6$, and $N_s = 3$.

## 4.5 Comparison with other datasets

Table 1 shows the metrics of applying PhyDNet on the three videos. It can be seen that the SSIM in the range of 0.7 can be achieved. While this is smaller than the SSIMs of applying PhyDNet on other standard dataset, which are in the range of 0.9 [6], it can be understood that our dataset only contain a few hundred video clips, and the number of video clips in standard dataset such as Moving MNIST is much larger [6]. In addition, our data is experimental data, which is less optimal than the data in the standard dataset, e.g., our TEM data contain noise. The SSIMs of greater than 0.7 using our dataset are still satisfied because the transition of nano-scale structure of the Au NP can already be captured by the model, as shown in Figure 3 and Figure 4. Hence, it is demonstrated that the PhyDNet model can give desirable performance on very limited scientific data. In the future work, we may use synthetic atomic-resolution data to train the PhyDNet model and use the trained model to predict the experimental data, a methodology adopted by many reported studies [15, 29, 30].

*Table 1: MAE, MSE, and SSIM of applying PhyDNet on the three Au NP videos.*

|      | 4-nm Au NP | <2-nm Au NP | SL Au NP |
|------|------------|-------------|----------|
| MAE  | 0.53       | 0.57        | 0.55     |
| MSE  | 0.46       | 0.51        | 0.51     |
| SSIM | **0.73**   | **0.76**    | **0.78** |

# 5 Conclusions

Deep learning models such as PhyDNet have achieved good performance on video prediction when the models were trained using standard benchmark datasets. In catalytic science, in-situ TEM videos have recorded the structure transition of Au NP from rigid to disordered dynamic structure, which might explain the superior catalytic properties of ultrasmall Au NP. Building on progress in these two fields, we demonstrate that the PhyDNet model can be applied to in-situ TEM videos for the prediction of structure evolution of gold nanoparticle, especially the transition from rigid to dynamic structure. The results show that PhyDNet can predict the structure transition of 2-nm and SL Au NPs from rigid to dynamic structure. The SSIMs measured between PhyDNet predicted and ground truth TEM images in the Au NP videos are in the range of 0.7. While these SSIMs are smaller than the values of about 0.9 when PhyDNet model was trained on the standard datasets, it is understandable because the amount of data from the Au NP videos is much smaller than that of the benchmark datasets. Our results indicate that PhyDNet is an effective deep learning model to provide desirable video prediction from limited scientific data. Because PhyDNet is a general architecture for video prediction, and video recording is a common method for material research, we anticipate that PhyDNet can be used on many other scientific videos not limited to the in-situ TEM videos, and the video-prediction capability we developed can be used for a variety of applications. In particular, it may find application in the development of automated experimentation and future data-driven microscope platforms [31, 32].

# 6 Acknowledgments

This research was supported by the Energy Storage Materials Initiative (ESMI), under the Laboratory Directed Research and Development (LDRD) Program at Pacific Northwest National Laboratory (PNNL). PNNL is a multi-program national laboratory operated for the U.S. Department of Energy (DOE) by Battelle Memorial Institute under Contract No. DE-AC05-76RL01830.

# 7  References

1. Shi, X., et al. *Convolutional LSTM network: A machine learning approach for precipitation nowcasting*. in *Advances in neural information processing systems*. 2015.
2. Kwon, Y.-H. and M.-G. Park. *Predicting future frames using retrospective cycle gan*. in *Proceedings of the IEEE/CVF Conference on Computer Vision and Pattern Recognition*. 2019.
3. Finn, C., I. Goodfellow, and S. Levine, *Unsupervised learning for physical interaction through video prediction.* Advances in neural information processing systems, 2016. **29**: p. 64-72.
4. LeCun, Y., Y. Bengio, and G. Hinton, *Deep learning.* Nature, 2015. **521**(7553): p. 436-444.
5. Wang, Y., et al. *Predrnn: Recurrent neural networks for predictive learning using spatiotemporal lstms*. in *Proceedings of the 31st International Conference on Neural Information Processing Systems*. 2017.
6. Guen, V.L. and N. Thome. *Disentangling physical dynamics from unknown factors for unsupervised video prediction*. in *Proceedings of the IEEE/CVF Conference on Computer Vision and Pattern Recognition*. 2020.
7. Srivastava, N., E. Mansimov, and R. Salakhudinov. *Unsupervised learning of video representations using LSTMs*. in *International conference on machine learning*. 2015. PMLR.
8. Yao, L., et al., *Machine learning to reveal nanoparticle dynamics from liquid-phase tem videos.* ACS central science, 2020. **6**(8): p. 1421-1430.
9. He, Y., et al., *Size-dependent dynamic structures of supported gold nanoparticles in CO oxidation reaction condition.* Proceedings of the National Academy of Sciences of the United States of America, 2018. **115**(30): p. 7700-7705.
10. Ge, M., et al., *Deep learning analysis on microscopic imaging in materials science.* Materials Today Nano, 2020. **11**.
11. Horwath, J.P., et al., *Understanding important features of deep learning models for segmentation of high-resolution transmission electron microscopy images.* npj Computational Materials, 2020. **6**(1): p. 1-9.
12. Akers, S., et al., *Rapid and flexible segmentation of electron microscopy data using few-shot machine learning.* npj Computational Materials, 2021. **7**(1): p. 187.
13. Wang, X., et al., *AutoDetect-mNP: an unsupervised machine learning algorithm for automated analysis of transmission electron microscope images of metal nanoparticles.* Jacs Au, 2021. **1**(3): p. 316-327.
14. Cho, P., et al., *Defect Detection in Atomic Resolution Transmission Electron Microscopy Images Using Machine Learning.* Mathematics, 2021. **9**(11): p. 1209.
15. Förster, G.D., et al., *A deep learning approach for determining the chiral indices of carbon nanotubes from high-resolution transmission electron microscopy images.* Carbon, 2020. **169**: p. 465-474.
16. Oprea, S., et al., *A review on deep learning techniques for video prediction.* IEEE Transactions on Pattern Analysis and Machine Intelligence, 2020.
17. Zhou, Y., H. Dong, and A. El Saddik, *Deep learning in next-frame prediction: a benchmark review.* IEEE Access, 2020. **8**: p. 69273-69283.

18. Jin, B., et al. *Exploring spatial-temporal multi-frequency analysis for high-fidelity and temporal-consistency video prediction.* in *Proceedings of the IEEE/CVF Conference on Computer Vision and Pattern Recognition.* 2020.
19. Chang, Z., et al., *MAU: A Motion-Aware Unit for Video Prediction and Beyond.* Advances in Neural Information Processing Systems, 2021. **34**.
20. Spurgeon, S.R., et al., *Towards data-driven next-generation transmission electron microscopy.* Nature Materials, 2021. **20**(3): p. 274-279.
21. Haruta, M., *Size-and support-dependency in the catalysis of gold.* Catalysis today, 1997. **36**(1): p. 153-166.
22. Haruta, M., et al., *Low-temperature oxidation of CO over gold supported on TiO2, α-Fe2O3, and Co3O4.* Journal of Catalysis, 1993. **144**(1): p. 175-192.
23. Herzing, A.A., et al., *Identification of active gold nanoclusters on iron oxide supports for CO oxidation.* Science, 2008. **321**(5894): p. 1331-1335.
24. Liu, Y., et al., *Highly active iron oxide supported gold catalysts for CO oxidation: how small must the gold nanoparticles be?* Angewandte Chemie, 2010. **122**(33): p. 5907-5911.
25. De Bézenac, E., A. Pajot, and P. Gallinari, *Deep learning for physical processes: Incorporating prior scientific knowledge.* Journal of Statistical Mechanics: Theory and Experiment, 2019. **2019**(12): p. 124009.
26. Ionescu, C., et al., *Human3. 6m: Large scale datasets and predictive methods for 3d human sensing in natural environments.* IEEE transactions on pattern analysis and machine intelligence, 2013. **36**(7): p. 1325-1339.
27. Gonzalez, R.C. and R.E. Woods, *Digital Image Processing.* 3rd ed. 2008: Prentice Hall.
28. Wang, Z., et al., *Image quality assessment: from error visibility to structural similarity.* IEEE transactions on image processing, 2004. **13**(4): p. 600-612.
29. Maxim, Z., et al., *Tracking atomic structure evolution during directed electron beam induced Si-atom motion in graphene via deep machine learning.* Nanotechnology, 2021. **32**(3).
30. Lin, R., et al., *TEMImageNet training library and AtomSegNet deep-learning models for high-precision atom segmentation, localization, denoising, and deblurring of atomic-resolution images.* Scientific reports, 2021. **11**(1): p. 1-15.
31. Olszta, M., et al., *An Automated Scanning Transmission Electron Microscope Guided by Sparse Data Analytics.* arXiv preprint arXiv:2109.14772, 2021.
32. Doty, C., et al., *Design of a graphical user interface for few-shot machine learning classification of electron microscopy data.* Computational Materials Science, 2022. **203**: p. 111121.
33. Shao M. et al. *Electrocatalysis on Platinum Nanoparticles: Particle Size Effect on Oxygen Reduction Reaction Activity.* Nano Letters 2011, 11, 9, 3714-3719

# Supplementary Information

# Deep-Learning-Based Prediction of Nanoparticle Phase Transitions During In Situ Transmission Electron Microscopy

**Authors:** Wenkai Fu[1], Steven R. Spurgeon[1,2], Chongmin Wang[1,§], Yuyan Shao[1], Wang Wei, Amra Peles[1,*]

**Affiliations:**
1. Pacific Northwest National Laboratory, 902 Battelle Blvd, Richland, WA 99354, USA
2. Department of Physics, University of Washington, Seattle, WA 98195, USA

\* Corresponding author. Email: amra.peles@ornl.gov

§ Corresponding author. Email: chongmin.wang@pnnl.gov

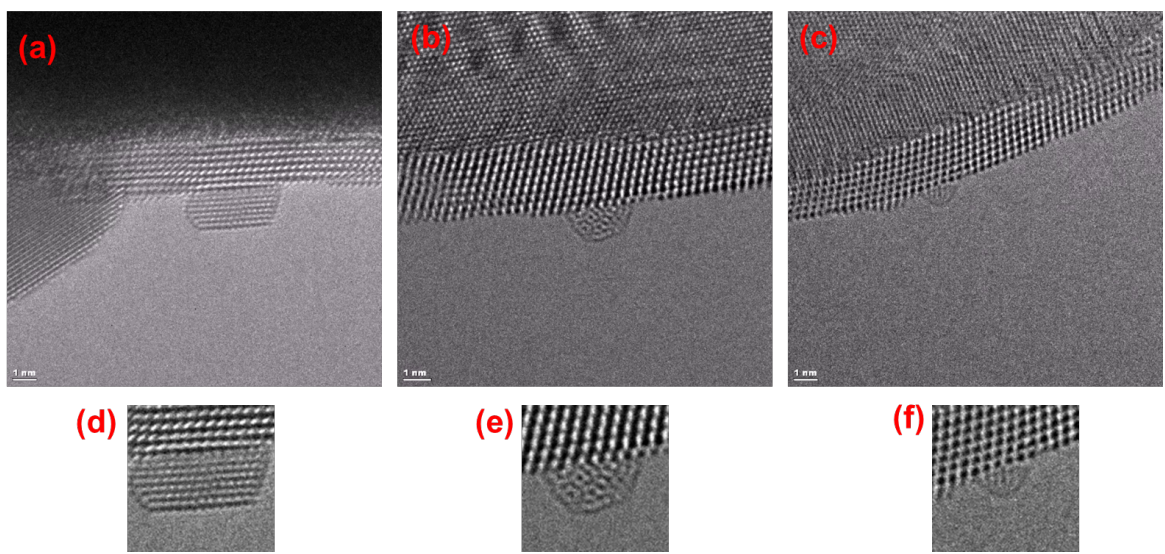

**Supplementary Figure S1**: *A 128x128 window is placed in the original frames to crop the region with Au NP centered. (a) to (c) are the original 480x480 frames from videos of 4-nm, 2-nm, and SL Au NP, respectively. (d) to (f) are their cropped and normalized frames with size 128x128.*

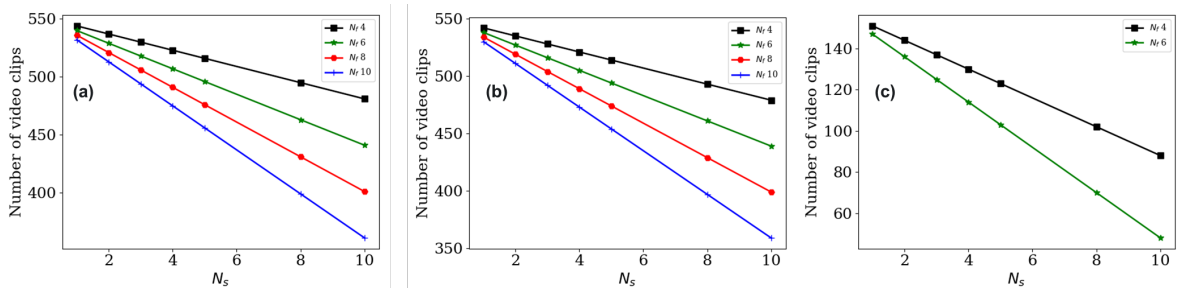

**Supplementary Figure S2**: *The resulted number of video clips as a function of NS and NF in the (a) 4-nm, (b) 2-nm, and (c) SL Au NP videos.*

# Effects of histogram matching algorithm

The histograms of the target, predicted, and histogram-matched frame 4 shown in Figure 3 are illustrated in Figure S3. Each of the grayscale frames has 128x128 pixels with intensities from 0 to 255. The predicted frame has two peaks at the bright (intensity = 0) and dark sides (intensity = 255), which correspond to the white and black regions inside the $CeO_2$ supporter. After histogram matching, the two peaks still exist, but they are shifted to the center region, i.e., intensities near 100 and 200. The target frame has small peaks between the intensities of 70 and 140. After histogram matching, histogram of the predicted frame also contains these peaks. Hence, the matched histogram is closer to the target one. The similarity of the target and the matched histograms is manifested by the similarity between the target and the histogram-matched frame 4 in Figure 3.

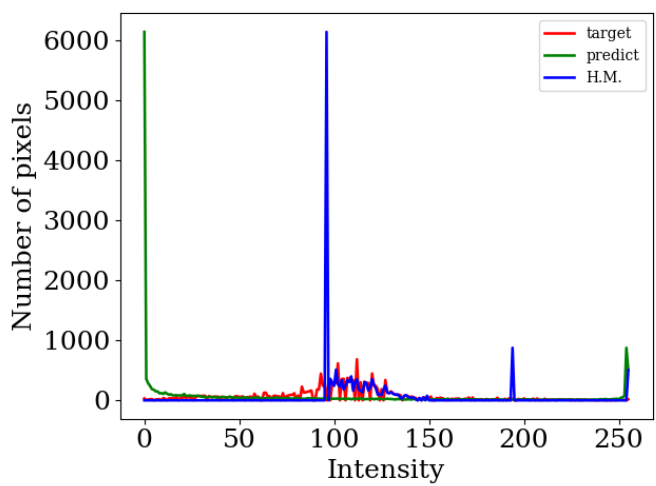

***Supplementary Figure S3***: *Histograms of the target, predict, and histogram matched frame 4 shown in Figure 5.*